# DC Characterization of a Circular, Epoxy-Impregnated High Temperature Superconducting (HTS) Coil

Di Hu, *Student Member, IEEE*, Mark D. Ainslie, *Member, IEEE*, Jordan P. Rush, John H. Durrell, and Jin Zou, *Student Member, IEEE*

*Abstract*—Direct current (DC) characterization of high temperature superconducting (HTS) coils is important for HTS applications, such as electric machines, superconducting magnetic energy storage (SMES) and transformers. In this paper, DC characterization of a circular, epoxy-impregnated HTS coil made from YBCO coated conductor for use as a prototype axial flux HTS electric machine is presented. Multiple voltage taps were utilized within the coil during measurement to help provide further detailed information on its DC behavior as a function of length. Based on the experimental results, there exist regions of non-uniformity along the length of superconductor in the coil, resulting in non-ideal superconducting properties of the coil. By studying the current-voltage (*I-V*) curves across different regions, it is found that a decreasing *n*-value and critical current exists in the non-uniform parts of the HTS coil.

## I. Introduction

RECENTLY, long lengths of high quality YBCO coated conductor have become commercially available, with increased potential for implementation in practical, large-scale high temperature superconducting (HTS) applications. HTS materials are able to carry much higher current densities than conventional materials, such as copper, and offer near zero resistance to direct current (DC) flow when cooled below a certain cryogenic temperature. Consequently, HTS coils are expected to be used in devices in order to improve the system's efficiency whilst reducing its system size and weight. The DC characterization of HTS coils is important for HTS device design, such as electrical machines, superconducting magnetic energy storage (SMES) and transformers. The DC properties determine the maximum current an HTS device may carry and helps determine the volume and power density of an electrical machine [1]-[3]. The DC characterization of HTS coils has been experimentally measured and simulated numerically by several groups [2]-[9]. However, few papers investigate the variation in its properties along the length of the tape using detailed DC characterization of the HTS coil. It is a common phenomenon that some non-uniformity exists along the length of an HTS coated conductor [2], [4], [5], [7], [10]-[14]. This variation may be for several reasons, such as an uneven manufacturing process or damage from cutting, handling or winding. The epoxy impregnation of HTS coils can also have a detrimental effect on its superconducting properties [2], [4], [13]. It is, therefore, important to consider the non-uniformity of the coated conductor along the length of the tape.

We are currently investigating an all-superconducting axial flux-type electric machine using HTS materials in both wire and bulk forms [15], [16] and carrying out the characterization of prototype HTS coils for this purpose. The circular, epoxy-impregnated HTS coil investigated in this paper is one such candidate for the stator winding of such a machine and is used in conventional axial flux-type machines. Based on experimental measurements, we investigate superconducting properties along the length of the coil and numerical analysis is carried out to help interpret the experiment results.

Voltages were measured at several points along the coil to provide detailed information. In section 2, the experimental details are described. The numerical analysis is performed using the *H*-formulation implemented in the commercial software package Comsol Multiphysics 4.3a [17]. In section 3, detailed results and discussion are presented. It is found that there are inhomogeneity/non-uniformity along the coil, resulting in an overall critical current significantly lower than expected.

## II. DC Characterization of the Test Coil

The circular, HTS pancake coil was wound using SuperPower's SCS4050-AP coated conductor [18] and then vacuum impregnated in epoxy and is shown in Fig. 1. The total length of the tape was 19.3 m, resulting in 47 layers of silk ribbon interleave. The vacuum impregnation process included a 10 bar overpressure of nitrogen gas maintained for around 5 hours before rotary baking. The resin used was stycast W19 with catalyst 11 in the ratio of 100:17. The specifications of the coated conductor and circular coil are shown in Table I.

There are ten voltage taps used within the coil to provide

The work of D. Hu and J. Zou were supported by Churchill College, the China Scholarship Council and the Cambridge Commonwealth, European and International Trust. The work of M. D. Ainslie was supported by a Royal Academy of Engineering Research Fellowship.

The authors are with the Bulk Superconductivity Group, Department of Engineering, University of Cambridge, Cambridge CB2 1PZ, U.K. (e-mail: dh455@cam.ac.uk).

more detailed information during its DC characterization. By pairing the voltage taps, there are 13 voltage sections for the specific measurements of the coil, as shown in Fig. 2. The voltages $V_1$-$V_9$ represent the voltages between the individual taps, which are spaced approximately every 2 m along the tape length. The first voltage tap is wired 1 m from the coil's inner copper current contact and the final voltage tap is wired 0.3 m from the outer current contact to prevent an anomalous measurement due to localized heating of the tape near the coil terminals [8, 16]. Hence, $V_1$ and $V_9$ represent the voltage across the innermost and outermost sections of the coil respectively. $V_{10}$, $V_{11}$, and $V_{12}$ represent the voltages along each third of the coil, where $V_{10} = V_1 + V_2 + V_3$, $V_{11} = V_4 + V_5 + V_6$, and $V_{12} = V_7 + V_8 + V_9$, and $V_{13}$ is total coil voltage across the entire coil.

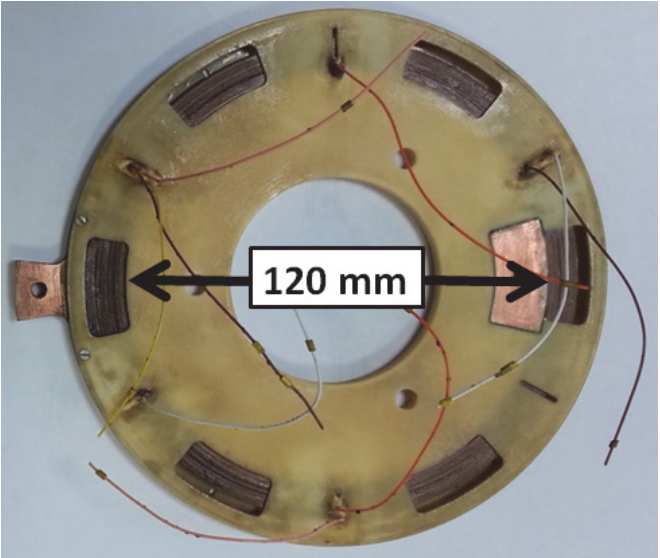

Fig. 1. The circular, HTS pancake coil used in this study.

TABLE I
SPECIFICATION OF THE COATED CONDUCTOR AND CIRCULAR, HTS PANCAKE COIL

| Parameter | Circular Pancake Coil |
|---|---|
| Tape manufacturer | SuperPower |
| Tape type | SCS4050-AP |
| Critical current [Self-field, 77 K] | 101.4 A |
| Conductor width, $w$ | 4 mm |
| Conductor thickness, $d_c$ | 0.1 mm |
| YBCO layer thickness, $d_{sc}$ | 1 µm |
| Distance between YBCO layers | Approximately 190 µm |
| Length of conductor used | 19.3 m |
| Coil turns, $n_c$ | 47 |
| Coil inner diameter | 120 mm |

The coil was cooled to 77 K over several hours by suspension in liquid nitrogen vapour followed by complete submersion in a liquid nitrogen bath. The current-voltage curves of the circular coil over the different voltage sections were recorded. The coil current was supplied by a DC power supply. The ramp-rate was set as low as 0.5 A/s to reduce the influence of any ac effects and the inductance of the coil. The critical current of the coil/voltage section region is defined as the DC current when the electric field across the region reaches the characteristic electric field, $E_0 = 100$ µV/m [19], [20], – this can be converted to a characteristic voltage using the distance between two taps. Although there can be a higher electric field (> 100 µV/m) in the inner turns of the coil [6], it is still the simplest way to define the critical current of the coil and can help us analyze the DC characteristics of the coil. The measurements were performed multiple times to verify reproducibility and to check the effect of thermal cycling between room temperature and 77K, which was insignificant.

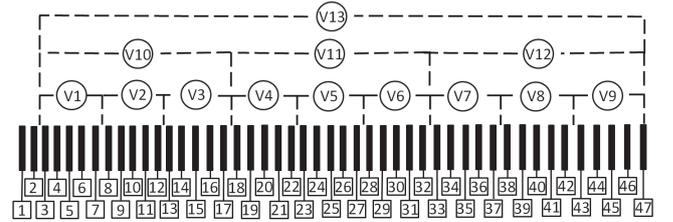

Fig. 2. The specific voltage taps and measured voltages for the circular, HTS pancake coil. Numbers at the bottom of the diagram represents the turn numbers of the coil.

A 2D axisymmetric model of the circular pancake coil [3], [6], [8] is set up to help study the ideal DC characteristics, based on the **H**-formulation [21]-[24] implemented in Comsol Multiphysics 4.3a [17]. A stack of tapes in the *rz*-plane is used in the model to represent the cross section of the coil. The inner radius of the coil to the innermost tape is 60 mm, the distance between each turn/tape is 190 µm, the width of the tape is 4 mm, and the thickness of the tape (i.e., the HTS layer) is 1 µm. Further details of the implementation of the **H**-formulation in Comsol Multiphysics can be found in [7], [25]-[27]. A short sample from the spool of tape used to wind the coil was measured in [28] and its measured angular dependence of the critical current density $J_c(B, \theta)$ is shown in Fig. 3. Following [7], [25]-[27], uniform critical current density is assumed along the length of tape in the circular coil for the numerical analysis.

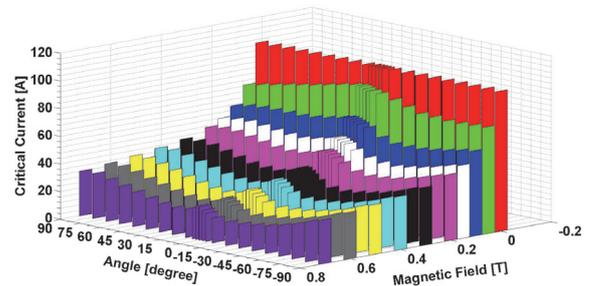

Fig. 3. The measured angular dependence of the critical current density, $J_c(B, \theta)$, of the short sample from the spool of tape.

During simulation, a two-variable direct interpolation is used to include the measured $J_c(B, \theta)$ [16], [28], as in Fig. 3. The applied current was ramped at 7 A/s. For HTS materials, *n*-value is usually within the range of 5 (strong flux creep) and 50 (limiting value for HTS and LTS materials) [29]-[31]. When the *n*-value is over 20, the result will be similar to Bean's critical state model [32]. In [33], the high performance of superconducting wire made by SuperPower is reported. The length of the wire is over a kilometer and the *n*-value is over 20 along the length of coated conductor. Therefore, we assume $n = 21$ in our simulation as a reasonable value.

The voltage across each section ($V_1$–$V_{13}$) is derived by integrating the local electrical field along its length.

The average electric field, and voltages $V_1$–$V_{13}$, are given by (1) and (2), respectively.

$$E_{ave,i}(i=1 \sim 13) = \sum_m \int 2\pi r E_\varphi \, ds \Big/ \sum_m \int 2\pi r \, ds \quad (1)$$

$$V_i(i=1 \sim 13) = E_{ave,i} l_i \quad (2)$$

where *i* denotes the voltage sections 1–13 shown in Fig. 2 and *m* denotes the turn number in each section (3 to 47 turns in the circular coil, corresponding to the positions of the voltage taps). $E_\varphi$ represents the local electric field in the $\varphi$-direction, *s* represents the cross-section of turn *m*, and $l_i$ represents the wire length of each voltage section. Taking voltage section 1 as an example, *m* varies from 3 to 7. The electric field and voltage can then be calculated by (1) and (2).

### III. RESULTS AND DISCUSSION

The current-voltage curves from the DC measurement and simulation results for the circular coil are shown in Fig. 4. $V_{13}$ represents the total coil voltage.

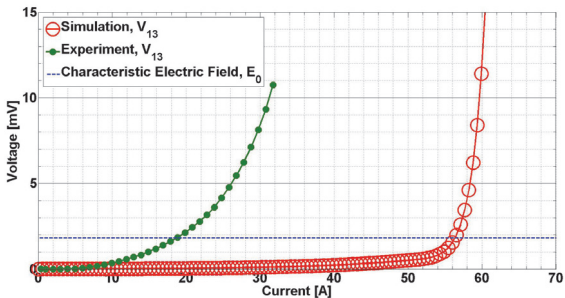

Fig. 4. The current-voltage curves from the experimental and simulation results of the circular coil.

Based on Fig. 4, the current-voltage curve from the experimental measurement is significantly different from the simulation results. The critical current of the coil is measured to be around 19 A, but the simulated critical current is around 56 A. By experiment, the *n*-value of the coil is only 3, whereas for the simulation, it was assumed the *n*-value of the coil is 21, as assumed in the model. Hence, the critical current and *n*-value obtained experimentally are much lower than the expected values in the simulation.

In order to investigate the possible reasons for this phenomenon, we investigate the uniformity of the coil based on the experimental data. The measured and simulated current-voltage curves for $V_{10}$–$V_{12}$ are shown in Fig. 5, where $V_{10}$, $V_{11}$, and $V_{12}$ are the curves for each third of the coil. In addition, the average electric field for voltage sections $V_1$–$V_9$ when the input current reaches the critical current of the whole coil (no higher to avoid quenching and potentially damaging the coil), and the test data for the critical current every 5 m along the coated conductor (provided by the manufacturer) are shown in Fig. 6.

Based on Figs. 5, 6(a) and 6(b), non-uniformity of the critical current exists along the length of the circular HTS coil. According to Fig 6(a), the maximum electric field, corresponding to the region of lowest critical current in the coil, exists between 13 m and 15 m (voltage $V_7$ in the circular coil). Based on Figure 6(b), there are regions of lowest critical current between 14.4 m and 19.3 m, indicating a potential overlap between the maximum electric field and lowest critical current. However, the scale of the fluctuations in our experimental results is not consistent with the scale of variation in $J_c$ as provided by the manufacturer.

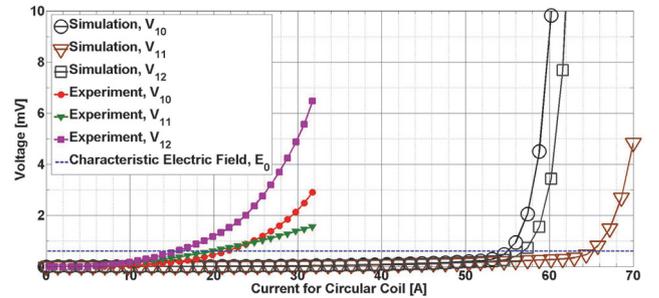

Fig. 5. Current-voltage curves for voltage section 10-12.

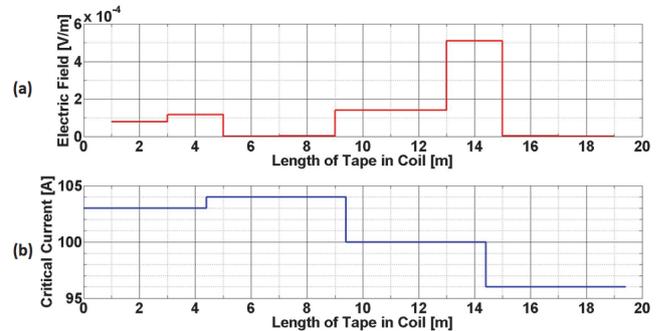

Fig. 6(a) Average electric field for voltage sections $V_1$–$V_9$, at a current of the critical current of the coil (b) Test data for the critical current every 5 m along the conductor, provided by the manufacturer.

Based on Fig. 5, it can be seen that the curves from the simulation and experiment are quite different. The critical current $I_c$ of each section can be determined by experiment and simulation tests. According to the experimental results, the $I_c$ ranking of these three sections is $I_c(V_{12}) < I_c(V_{11}) < I_c(V_{10})$. Based on the simulation, the $I_c$ ranking of these three sections is $I_c(V_{10}) < I_c(V_{12}) < I_c(V_{11})$. Uniform critical current density is assumed along the length of the tape in the simulation as previous stated. In the ideal case, it is expected that the inner turns, i.e., $V_{10}$, will see a relatively higher localised magnetic field, which would determine the overall critical current of the coil [6], [8], [9]. Therefore, the difference between the experiment and simulation shows that the coil is not uniform along its length.

Based on Fig. 6(a), there are significant electric fields generated in voltage sections $V_1$, $V_2$, $V_5$, $V_6$, and $V_7$, which dominate the behavior of the whole coil. These are identified as the "non-uniform" regions. For the other regions, there are comparatively lower electric fields, and these are identified as the "uniform" regions.

From Fig. 6(b), there is non-uniform critical current along the length of tape before epoxy impregnation. Using the information in Fig. 6(b), a simulation is carried out and compared with the experiment results, which is shown in Fig. 7.

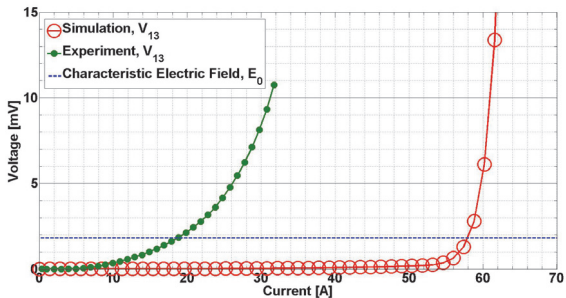

Fig. 7. Current-voltage curves for the experimental and simulation results of the circular HTS coil after considering the critical current data along the length shown in Fig. 6(b).

Based on Fig. 7, the critical current of the coil based on the new simulation is 58 A, which is only slightly different from the previous simulation result. There is still a large difference between the simulation and experimental results. This indicates that besides material processing, some other reason, such as the epoxy impregnation, causes the lower critical current of the coil. In [2], [14], it is reported that mechanical stress can cause a decreased $n$-value and critical current in an impregnated coil due to the difference in thermal contraction of the materials in the coil, such as the coil former, conductor constituents, impregnation resin, bandages and heat-sink materials. Therefore, the impregnation could be considered the most important reason for the decreasing of $n$-value and critical current of the coil in this case.

To provide more information, the $n$-value and critical current in the non-uniform regions of the coil are analysed. It is found that in some non-uniform regions, the $n$-value cannot be considered as constant and varies when the magnitude of the current $I$ is changed. The specific details are shown in Table II.

For the uniform regions ($V_3$, $V_4$, $V_8$, $V_9$), there is almost no electric field when the input current is 32 A. It can be estimated from this that the critical current in these uniform regions is higher than 32 A. It is dangerous to further increase the input current in the experiment as a higher input current may lead to localized quenching and damage to the coil from heat generated in the regions of lowest critical current, such as $V_7$.

TABLE II
COMPARISONS OF N VALUE AND CRITICAL CURRENT IN NON-UNIFORM REGIONS

| Region | $N$-Value | Critical Current |
|---|---|---|
| $V_1$ | 4.03 | 20 A |
| $V_2$ | 2.96 ($I <$ 18 A) | 18 A |
|  | 2.4 ($I >$ 18 A) |  |
| $V_5$ | 2.44 ($I <$ 17 A) | 17 A |
|  | 1.98 ($I >$ 17 A) |  |
| $V_6$ | 2.44 ($I <$ 17 A) | 17 A |
|  | 1.98 ($I >$ 17 A) |  |
| $V_7$ | 3.8 ($I <$ 11 A) | 11 A |
|  | 2.83 ($I >$ 11 A) |  |

Based on Table II, the $n$-value in the non-uniform regions is much smaller than 20, and the critical current in these regions is much smaller than 32 A, which is far from ideal, given the potential of the tape. Based on [6], [8], [9], the critical current of the coil should be determined by inner turns of the coil if the coil is uniform along the length, i.e., the region with highest localized magnetic field. The non-uniform region with the lowest critical current (see Table II), i.e., region $V_7$, has the most significant influence on the critical current of the coil. This highlights the importance of taking detailed DC measurements along the coil's length.

## IV. CONCLUSION

In this paper, the DC properties of a circular, epoxy-impregnated HTS coil made from YBCO coated conductor for a prototype axial flux HTS electric machine were measured. Multiple voltage taps were used in the coil for the measurement to provide further detailed information on the DC characterization of the coil. Numerical analysis is carried out using a 2D axisymmetric finite element model to help analyse the experiment result.

Based on a comparison of the experimental and simulation results, there is a significantly reduced critical current and $n$-value in the coil. By further investigation, there are regions of suppressed $I_c$ and $n$-value present in the coil. It is thought that the material processing of the coated conductor (by the manufacturer) and subsequent coil manufacturing both cause non-uniformity in the coil, with the latter having a more significant effect on the coil performance.

By evaluating the electric field when the input current reaches the critical current of the whole coil, there are uniform and non-uniform regions in the coil. From the current-voltage curves, the critical currents in the non-uniform regions were found to be much lower than those in the uniform regions. The $n$-value in the non-uniform regions is much lower than 20. Because of these variations, the inner turns cannot simply be considered to determine the critical current of the coil as expected by simulation. The region with lowest critical current could be found from detailed measurements using internal voltage taps.